\documentclass[showkeys]{revtex4}
\usepackage{graphicx}
\usepackage{dcolumn}
\usepackage{amsmath}
\usepackage{color}
\usepackage{epstopdf}
\usepackage{float}
\usepackage{graphicx}
\usepackage{subfigure}
\usepackage[utf8x]{inputenc}
\definecolor{coolblack}{rgb}{0.0, 0.18, 0.39}
\definecolor{darkred}{rgb}{0.5,0,0}
\definecolor{darkgreen}{rgb}{0,0.5,0}
\definecolor{darkblue}{rgb}{0,0,0.5}
\definecolor{lapislazuli}{rgb}{0.15, 0.38, 0.61}
\definecolor{venetianred}{rgb}{0.78, 0.03, 0.08}
\definecolor{bleudefrance}{rgb}{0.19, 0.55, 0.91}
\definecolor{dogwoodrose}{rgb}{0.84, 0.09, 0.41}
\usepackage[linktocpage,colorlinks]{hyperref}
\hypersetup{colorlinks=true, citecolor=darkgreen, linkcolor=blue,urlcolor = darkblue}
\makeatletter
\def\btt#1{\texttt{\@backslashchar#1}}
\DeclareRobustCommand\bblash{\btt{\@backslashchar}} \makeatother

\RequirePackage{fix-cm}

\begin{document}
\title{Stability analysis of circular orbits around a traversable wormhole with massless conformally coupled scalar field %\thanksref{t1}  
}
\author{Shobhit Giri $^{a}$}\email{shobhit6794@gmail.com}
\author{ Hemwati Nandan $^{a,b}$}\email{hnandan@associates.iucaa.in}
\author{Lokesh Kumar Joshi $^{c}$}\email{lokesh.joshe@gmail.com}
\author{Sunil D. Maharaj $^{d}$}\email{maharaj@ukzn.ac.za}
\affiliation{$^{a}$Department of Physics, Gurukula Kangri (Deemed to be University), Haridwar 249 404, Uttarakhand, India}
\affiliation{$^{b}$Center for Space Research, North-West University, Mahikeng 2745, South Africa}
\affiliation{$^{c}$ Department of Applied Science, Faculty of Engineering and Technology, Gurukula Kangri (Deemed to be University), Haridwar 249 404, Uttarakhand, India}
\affiliation{$^{d}$ Astrophysics Research Centre, School of Mathematics, Statastics and Computer Science, University of KwaZulu-Natal, Private Bag X54001, Durban 4000, South Africa }


\begin{abstract}
\noindent  We study the stability of circular orbits in the background of a traversable wormhole (TWH) spacetime obtained as a solution of Einstein’s field equations coupled conformally to a massless scalar field. The  Lyapunov stability approach is employed to determine the stability of circular orbits (timelike and null) of non-spinning test particles around a TWH spacetime. In the case of timelike geodesics, the particle is confined to move in four different types of effective potentials depending on various values of the angular momentum $\tilde{L}$ with both centrifugal and gravitational part. The effective potential for null geodesics consists of only a centrifugal part. Further, we characterize each fixed point according to its Lyapunov stability, and thus classify the circular orbits at the fixed point into stable center and unstable saddle points by depicting the corresponding phase-portraits.
% \PACS{0.4.70.-s \and 04.20.-q \and 04.20.Cv \and 97.60.Lf}
\end{abstract}
\maketitle
\section{Introduction}
In view of general theory relativity (GR), Einstein and Rosen predicted the existence of bridges connecting two distant regions in spacetime named as Einstein-Rosen bridges or wormholes (WHs) \cite{einstein1935particle,visser1995lorentzian}.  Thereafter, the pioneering work of Morris and Thorne \cite{morris1988wormholes} developed the topological connections between separated regions of spacetime, as solutions of Einstein's field equations in GR, leading to the  formation of three-dimensional TWH geometries. Such WHs are considered as static and spherically symmetric spacetimes connecting two asymptotically flat regions with different physical properties. In the last few decades, a variety of studies in various aspects including gravitational lensing by wormholes \cite{cramer1995natural,perlick2004exact,safonova2001microlensing,kuhfittig2014gravitational}, shadows cast by WHs \cite{ohgami2015wormhole,nedkova2013shadow,shaikh2018shadows}, accretion disks surrounding them \cite{bambi2013broad,harko2008electromagnetic,harko2009thin}, and their viability as black hole alternatives \cite{damour2007wormholes,cardoso2016erratum,bueno2018echoes} have been widely investigated in detail. The violation of the null energy condition (NEC) \cite{hawking1973large,lobo2017wormholes} in the stress-energy tensor is required for construction of TWH which can be achieved by an exotic matter source responsible for the respective geometry \cite{barcelo1999traversable,willenborg2018geodesic,bohmer2008wormhole,flanagan1996does,ellis1979evolving,kleihaus2014rotating,chew2016geometry}. In fact, there is no need of any exotic matter in order to obtain TWH spacetimes in many modified theories of gravity \cite{fukutaka1989wormholes,lobo2009wormhole,kanti2011wormholes,harko2013modified}. In order to avoid the presence of an exotic matter in such geometries, Barcelo and Visser \cite{barcelo1999traversable} examined the conformal coupling of massless scalar field with gravity providing a large class of TWHs which also violate the NEC. They obtained a three-parameter class of exact solutions including the Schwarzschild geometry, certain naked singularities, and a collection of TWHs as analytical solutions of Einstein's field  equations. In view of such a conformal coupling, Callan et al. \cite{callan1970new}  introduced a “new improved stress-energy tensor” to obtain an interesting set of TWHs.\\
The new form of energy-momentum tensor which is also capable  of violating the NEC for a massless conformal scalar field $\phi_{c}$, is defined as
\begin{equation}
T_{\alpha \beta} = \nabla_{\alpha}\phi_{c} \nabla_{\beta}\phi_{c}-\frac{1}{2} g_{\alpha \beta} \left(\nabla\phi_{c}\right)^{2}+\frac{1}{6}\Big[G_{\alpha\beta}\phi_{c}^{2} -2\nabla_{\alpha}\left(\phi_{c}\nabla_{\beta}\phi_{c}\right)+2g_{\alpha\beta}\nabla^{\mu}\left(\phi_{c}\nabla_{\mu}\phi_{c}\right)\Bigr],
\end{equation}
where, $G_{\alpha\beta}$ and $g_{\alpha \beta}$ represents Einstein's tensor and metric tensor of spacetime respectively.\\

\noindent In this article, we wish to focus attention on static and spherically symmetric TWH geometries which were obtained by Barcelo and Visser \cite{barcelo1999traversable} by relating them conformally to the Janis–Newman–Winicour–Wyman (JNWW) solution \cite{janis1968reality,wyman1981static,virbhadra1997janis}. The scalar field corresponding to JNWW solutions is represented by
\begin{equation}
\phi_{m}=\sqrt{\frac{\left(8\pi G_{N}\right)^{-1}}{2}}\sin\chi ~\text{ln} \left(1-\frac{2\eta}{r}\right),
\end{equation}
where $G_{N}$ is Newton's constant.
In spherical polar coordinates ($t, r, \theta, \phi $), Agnese et al. \cite{agnese1985gravitation} obtained the JNWW metric as solutions of the Einstein field equations,  minimally coupled to a massless scalar field $\phi_{m}$, as given below
\begin{equation}
ds_{m}^{2}= -\left(1-\frac{2\eta}{r}\right)^{\cos\chi}dt^{2}+\left(1-\frac{2\eta}{r}\right)^{-\cos\chi}dr^{2}+\left(1-\frac{2\eta}{r}\right)^{1-\cos\chi} r^{2} \left(d\theta^{2}+\sin^{2}\theta d\phi^{2}\right),
\end{equation}
where parameters $\chi\in [0,\pi]$ and $\eta\geq0$.\\

\noindent Considering the new improved energy-momentum tensor $T_{\alpha\beta}$, Barcelo and Visser \cite{barcelo1999traversable} solved the Einstein's field equations
\begin{equation}
G_{\alpha\beta}= \left(8\pi G_{N}\right) T_{\alpha\beta},
\end{equation}
with $G_{\alpha\beta} =R_{\alpha\beta}-\frac{1}{2}g_{\alpha\beta}R$.
\noindent Since $T_{\alpha\beta}$ is traceless for a conformal field i.e. $T_{\alpha\beta}~g^{\alpha\beta}=0$, then the scalar curvature $R=0$. Thus the Einstein field equations coupled to conformal scalar field $\phi_{c}$ can be written as  \cite{barcelo1999traversable}
\begin{equation}
R_{\alpha\beta}=\left(\kappa-\frac{1}{6}\phi_{c}^{2}\right)^{-1}\left(\frac{2}{3}\nabla_{\alpha}\phi_{c}\nabla_{\beta}\phi_{c}-\frac{1}{6}g_{\alpha\beta}(\nabla\phi_{c})^{2}\right.
\left.-\frac{1}{3}\phi_{c}\nabla_{\alpha}\nabla_{\beta}\phi_{c}\right),
\end{equation}
\begin{equation}
\nabla^{2}\phi_{c}=0.
\end{equation}

\noindent Any metric $ds$ conformal to the JNWW metric $ds_{m}$ with conformal factor $\Omega(r)$ requires \cite{willenborg2018geodesic}
\begin{equation}
ds=\Omega(r) ds_{m}.
\end{equation}
The conformal factor $\Omega(r)$ with two real constants $\gamma_{+}$ and $\gamma_{-}$ can be expressed as \cite{barcelo1999traversable}
\begin{equation}
\Omega(r) = \gamma_{+} \left(1-\frac{2\eta}{r}\right)^{\frac{\sin\chi}{2\sqrt{3}}}+\gamma_{-} \left(1-\frac{2\eta}{r}\right)^{-\frac{\sin\chi}{2\sqrt{3}}}.
\end{equation}
We have a three-parameter set $(\chi, \eta, \Delta)$,  by defining an angle $\Delta$, such that
\begin{equation}
\tan\frac{\Delta}{2}=  \frac{\gamma_{+}~-~\gamma_{-}}{\gamma_{+}~+~\gamma_{-}} =  \frac{\bar{\gamma_{+}}~-~1}{\bar{\gamma_{+}}~+~1},
\end{equation} 
having range $ \Delta \in (-\pi,\pi] $ and $\bar{\gamma_{+}}= \frac{\gamma_{+}}{\gamma_{-}}$. \\

\noindent One obtains various spacetime metrics depending on the choice of these
parameters (i.e $\chi, \eta, \Delta$). The Schwarzschild black hole (SBH) metric can be recovered from $ds^{2}$ in the prescribed limit $\chi=0$ and for arbitrary values of $\eta$ and $\Delta$.\\

\noindent We intend to investigate the stability of circular orbits of test particles in the vicinity of TWH spacetime conformally coupled to a massless scalar field by using Lyapunov (in)stability criteria (as discussed in \cite{abolghasem2013stability} and references therein). Further, by analysing the phase-portrait in both cases of circular geodesics, we characterize each circular orbit at the fixed point as a stable center or an unstable saddle point. Boehmer et al. \cite{boehmer2012jacobi} extensively described two methods namely Lyapunov stability analysis \cite{abolghasem2012liapunov} and Jacobi stability analysis \cite{abolghasem2012jacobi}, or the Kosambi-Cartan-Chern (KCC) theory \cite{kosambi2016parallelism,cartan2016observations,key0000889m} for analysing stability of a dynamical system which plays crucial role in gravitation and cosmology.
The behavior of steady states of any dynamical system is analyzed to investigate the stability of the system. Lyapunov linear stability analysis requires the linearization of the dynamical system using the Jacobian matrix of a nonlinear system. However, both the linear and nonlinear stability analysis can be performed via Lyapunov stability approach \cite{abolghasem2012liapunov,abolghasem2013stability}. Besides, the Jacobi stability or the KCC theory is formulated in terms of a deviation equation of a second-order differential equation defining the whole trajectory subject to small perturbations. The trajectories of the structure equations those bunch together are Jacobi stable while trajectories are Jacobi unstable if they disperse on approaching the fixed point \cite{boehmer2012jacobi}.
In comparison to Jacobi stability analysis, Lyapunov approach is therefore more convenient to study the stability analysis as the linearization of nonlinear system takes place around a fixed point so-called equilibrium point.\\ 

\noindent Both methods of stability analysis are extensively discussed and implemented to illustrate the stability of circular orbits in the vicinity of SBH spacetime by Hossein \cite{abolghasem2013stability}. However, the Jacobi stability of dynamical systems employing KCC theory has already been analyzed in diverse situations \cite{balan2009berwald,boehmer2010nonlinear,harko2008jacobi,sabau2005some,boehmer2012jacobi}. We shall adopt the Lyapunov stability criteria to analyse the stability in our specified background geometry. 
A mathematical treatment of the concept of Lyapunov stability for a dynamical system has been discussed previously (see section-2 in \cite{abolghasem2013stability}).

\section{Geodesics around TWH geometry with massless conformally coupled scalar field }
This section is devoted to study the particle motion around a TWH spacetime obtained for an appropriate choice of parameters $\left(\chi=\frac{\pi}{3}, \Delta \neq \frac{\pi}{2} \right)$. Since a three parameter family of solution space is invariant under the transformation $(\eta,\chi, \Delta) \rightarrow (\eta, -\chi, -\Delta)$. Under transformation $\chi \rightarrow -\chi$, the JNWW solutions carry obvious symmetries with $\phi_{m}=-\phi_{m}$. However, for the transformation $(\eta,\chi, \Delta) \rightarrow (-\eta, \chi+\pi, \Delta)$, an additional symmetry is obtained under the coordinate transformation $r \rightarrow \tilde{r}= r-2\eta$ (with $\phi_{m}=+\phi_{m}$). Therefore, one can consider $\eta\geq 0$ and $\chi= [0,\pi]$ without loss of generality in view of these two symmetries. Nevertheless, only those geometries are examined in which $\Delta \neq \pi$ because they do not possess an asymptotic flat region as $r\rightarrow \infty$. Furthermore, the behavior of Ricci tensor component $R_{tt}$ as $r$ approaches $2\eta$ shows the divergence of geometry except for the parameter values 
$\chi=0$, $\{\chi=\frac{\pi}{3}, \Delta\neq \frac{\pi}{2}\}$, $\{\chi=\frac{2\pi}{3}, \Delta = \frac{\pi}{2}\}$ and 	$\chi=\pi$. The geometries are found with a naked curvature singularity at $r=2\eta$ for other than these parameter sets \cite{barcelo1999traversable,willenborg2018geodesic}. Therefore, for our choice of parameters $\{\chi=\frac{\pi}{3}, \Delta\neq \frac{\pi}{2}\}$, the geometry can be extended beyond $r=2\eta$ which reveals the existence of a genuine wormhole solution.\\

\noindent It is convenient to transform the radial coordinate $r$ into a new isotropic radial coordinate $\bar{r}$ by the relation \cite{barcelo1999traversable,willenborg2018geodesic}
\begin{equation}
r= \bar{r} \left(1+\frac{\eta}{2 \bar{r}}\right)^{2},
\end{equation}
where $\bar{r}$ ranges from $\frac{\eta}{2}$ to $\infty$.
Thus the metric conformal to the JNWW metric of TWH, supported by a massless conformally coupled scalar field, is then represented as \cite{barcelo1999traversable}
\begin{equation}
ds^{2}= \left[\gamma_{+}\left(\frac{1-\frac{\eta}{2\bar{r}}}{1+\frac{\eta}{2\bar{r}}}\right)+\gamma_{-}\right]^{2} \Bigl[-dt^{2}+\left(1+\frac{\eta}{2\bar{r}}\right)^4 
 \left[ d\bar{r}^{2}+ \bar{r}^{2} \left(d\theta^{2}+\sin^{2}\theta d\phi^{2}\right)\right]\Bigr].
\end{equation}

\noindent Now we investigate the geodesic motion of test particles in the equatorial plane of the TWH spacetime using Lagrangian approach. The Lagrangian for the motion (setting $\theta=\frac{\pi}{2}$) is given by
\begin{equation}
2\mathcal{L}= \left[\gamma_{+}\left(\frac{1-\frac{\eta}{2\bar{r}}}{1+\frac{\eta}{2\bar{r}}}\right)+\gamma_{-}\right]^{2} \left[-\dot{t}^{2}+\left(1+\frac{\eta}{2\bar{r}}\right)^4 \left( \dot{\bar{r}}^{2}+ \bar{r}^{2} \dot{\phi}^{2}\right)\right],
\end{equation}
where, a overdot represents differentiation w.r.t. the affine parameter $\tau$.
The spacetime metric admits time translational symmetry (i.e. independent of $t$) and rotational symmetry (i.e. independent of $\phi$) due to which corresponding generalized momenta are constants of motion. Using the Euler-Lagrange equations of motion, we deduce the first integral of geodesics equation for $t$ and $\phi$ by introducing two conserved quantities energy $(E)$ and angular momentum $(L)$, measured for unit mass of a test particle as

\begin{equation}
\dot{t}= \frac{E}{\left[\gamma_{+}\left(\frac{1-\frac{\eta}{2\bar{r}}}{1+\frac{\eta}{2\bar{r}}}\right)+\gamma_{-}\right]^{2}},\label{tdot}
\end{equation}

\begin{equation}
\dot{\phi}= \frac{L}{\left[\gamma_{+}\left(\frac{1-\frac{\eta}{2\bar{r}}}{1+\frac{\eta}{2\bar{r}}}\right)+\gamma_{-}\right]^{2}\left(1+\frac{\eta}{2\bar{r}}\right)^4 \bar{r}^{2}}.\label{phidot}
\end{equation}
\noindent For convenience, we define two quantities as
\begin{equation}
\left(\gamma_{+}\left(\frac{1-\frac{\eta}{2\bar{r}}}{1+\frac{\eta}{2\bar{r}}}\right)+\gamma_{-}\right) = \Theta , \label{Theta}
\end{equation}
\begin{equation}
\left(1+\frac{\eta}{2\bar{r}}\right) = \Pi . \label{Pi}
\end{equation}
The geodesics of test particles are constrained by the equation 
\begin{equation}
\Theta^{2} \left[-\dot{t}^{2}+\Pi^4\left( \dot{\bar{r}}^{2}+ \bar{r}^{2} \dot{\phi}^{2}\right)\right]= k. \label{constraint}
\end{equation}
One can obtain timelike and null geodesics for $k=-1$ and $k=0$  respectively with metric signature $(- + + +)$.
On substituting Eqs.\eqref{tdot} and \eqref{phidot} into constraint Eq.\eqref{constraint}, the radial equation for geodesics in the specified spacetime is derived as
\begin{equation}
\Theta^{4} ~\Pi^{4}~\frac{\dot{\bar{r}}^{2}}{2} = \frac{E^{2}}{2} - \frac{1}{2}\left(-k~ \Theta^{2} +\frac{L^{2}}{\bar{r}^{2}~ \Pi^{4}} \right)\label{radialeq}.
\end{equation}
The radial motion of the test particle on geodesics represented by Eq. \eqref{radialeq} is considered to be motion of a particle (with energy $E^{2}/2$) in the context of Newtonian mechanics moving in the effective potential $V_{eff}$. Therefore one can identify the above equation as a Newtonian central force problem  which reads
\begin{equation}
\frac{\dot{\bar{r}}^{2}}{2}= \tilde{E}- V_{eff}(\bar{r}). \label{cfe}
\end{equation}
The derivative of Eq.(\eqref{cfe}) with respect to affine parameter $\tau$ gives
\begin{equation}
\ddot{\bar{r}}= -V_{eff}'(\bar{r}). \label{1dradial}
\end{equation}
Here and throughout the work, $(')$ denotes differentiation w.r.t. radial coordinate.\\

\noindent Further, we are looking for Lyapunov stability analysis of geodesics by transforming the one-dimensional radial Eq.(\eqref{1dradial})  to a first order differential equation system in $\bar{r}-p$ phase-space as follows
\begin{equation}
\dot{\bar{r}}= p~,
~~~~~~ \dot{p}= -V_{eff}'(\bar{r}). \label{system}
\end{equation}
Hence Eq.(\eqref{cfe}) reduces to the following form
\begin{equation}
\frac{{p}^{2}}{2} + V_{eff}(\bar{r})= \tilde{E}.\label{eq22}
\end{equation}
Now we consider a vector field for the system of Eqs.\eqref{system} in order to linearize the above Eq.\ref{eq22} as
\begin{equation}
F(\bar{r},p) =\left(p, V_{eff}'(\bar{r})\right).
\end{equation}
At any point $(\bar{r}, p)$, one can find  the Jacobian matrix of $F$ as
\begin{equation}
J= \frac{\partial F(\bar{r},p)}{\partial q} = \begin{bmatrix}
0 & 1\\
-V_{eff}''(\bar{r}) &  0
\end{bmatrix}, 
\end{equation}
where, $q$ is a generalized coordinate.
The  characteristic equation $|J-\lambda I|=0$ yields the eigenvalues of Jacobian $J$ as follows
\begin{equation} \lambda=\pm \sqrt{-V_{eff}''(\bar{r})} . \label{eigen}
\end{equation}
Assuming a equilibrium point or a fixed point $(\bar{r}_{0}, 0)$, the  $V_{eff}'(\bar{r_{0}})$ vanishes, and for which $V_{eff}''(\bar{r_{0}})\neq 0 $. We can characterize the fixed point  $(\bar{r}_{0}, 0)$ as a saddle point corresponding to $V_{eff}''(\bar{r_{0}}) < 0 $ for which $\lambda$ is real and a possible center corresponding to $V_{eff}''(\bar{r_{0}}) > 0 $ for which $\lambda$ is imaginary. In other words, we can say that when the potential has a local maximum at $\bar{r}_{0}$ then $(\bar{r}_{0}, 0)$ is said to be a saddle point, and when the potential has local minimum at $\bar{r}_{0}$ then $(\bar{r}_{0}, 0)$ is said to be a possible center.
In particular, the energy function \eqref{eq22} can be interpreted as a Lyapunov function $\tilde{E}(\bar{r},p)$. The other required condition on the Lyapunov function is  that there should be a local minimum of $\tilde{E}$ at the fixed point for which the Hessian matrix can be written as \cite{abolghasem2013stability}
\begin{equation}
H_{\tilde{E}}= \frac{\partial^{2} \tilde{E}(\bar{r},p)}{\partial \bar{r} \partial p} = \begin{bmatrix}
V_{eff}''(\bar{r}) & 0\\
0 &  1
\end{bmatrix}. 
\end{equation}
Indeed, the fixed point  $(\bar{r}_{0}, 0)$ can be identified as a possible center as the above matrix is positive definite for  $V_{eff}''(\bar{r_{0}}) > 0 $. Therefore, the Lyapunov function has a local minimum at that point. Interestingly, the fixed point $(\bar{r}_{0}, 0)$ is Lyapunov stable for $V_{eff}''(\bar{r_{0}}) > 0 $ and Lyapunov unstable for $V_{eff}''(\bar{r_{0}}) < 0 $.\\

\noindent In order to classify the possible orbits of test particles around TWH spacetime, we need two coordinate transformations (see \cite{willenborg2018geodesic} for details) by introducing a radial coordinates as follows
\begin{equation}
\hat{x}= \frac{\bar{r}}{\eta}+\frac{1}{2},\label{tr1}
\end{equation}
and then transform $\hat{x}$ by 
\begin{equation}
\hat{r}= \frac{1}{2 \hat{x}}.\label{tr2}
\end{equation}
The new radial coordinate $\hat{r}\in [0,1]$ and the location of the throat of TWH is then given by \cite{willenborg2018geodesic}
\begin{equation}
\hat{r}_{T}= \frac{1}{\sqrt{\arrowvert (\bar{\gamma_{+}}-1)/(\bar{\gamma_{+}}+1)}\arrowvert+1}.
\end{equation}
On substituting the scaled quantities $\bar{\gamma_{+}}=\gamma_{+}/\gamma_{-}$ and $\tilde{L}=L/\eta \gamma_{-}$ with transformation Eqs.\eqref{tr1} and \eqref{tr2} into $V_{eff}$ (i.e. Eq.\eqref{cfe}), the effective potential in terms of the new radial coordinate $\hat{r}$ takes the following form
\begin{equation}
V_{eff}(\hat{r})=  2 \tilde{L}^{2} \hat{r}^{2}(1-\hat{r})^{2}-\frac{k}{2} [\bar{\gamma_{+}}(2\hat{r}-1)-1]^{2}. \label{effF}
\end{equation}
Hence the effective potential for TWH spacetime, conformally coupled to massless scalar field, has a centrifugal part and a gravitational part represented by the first and second terms of the above equation respectively. 
\begin{figure*}[]
	\centering
	\subfigure[]{\includegraphics[scale=0.7]{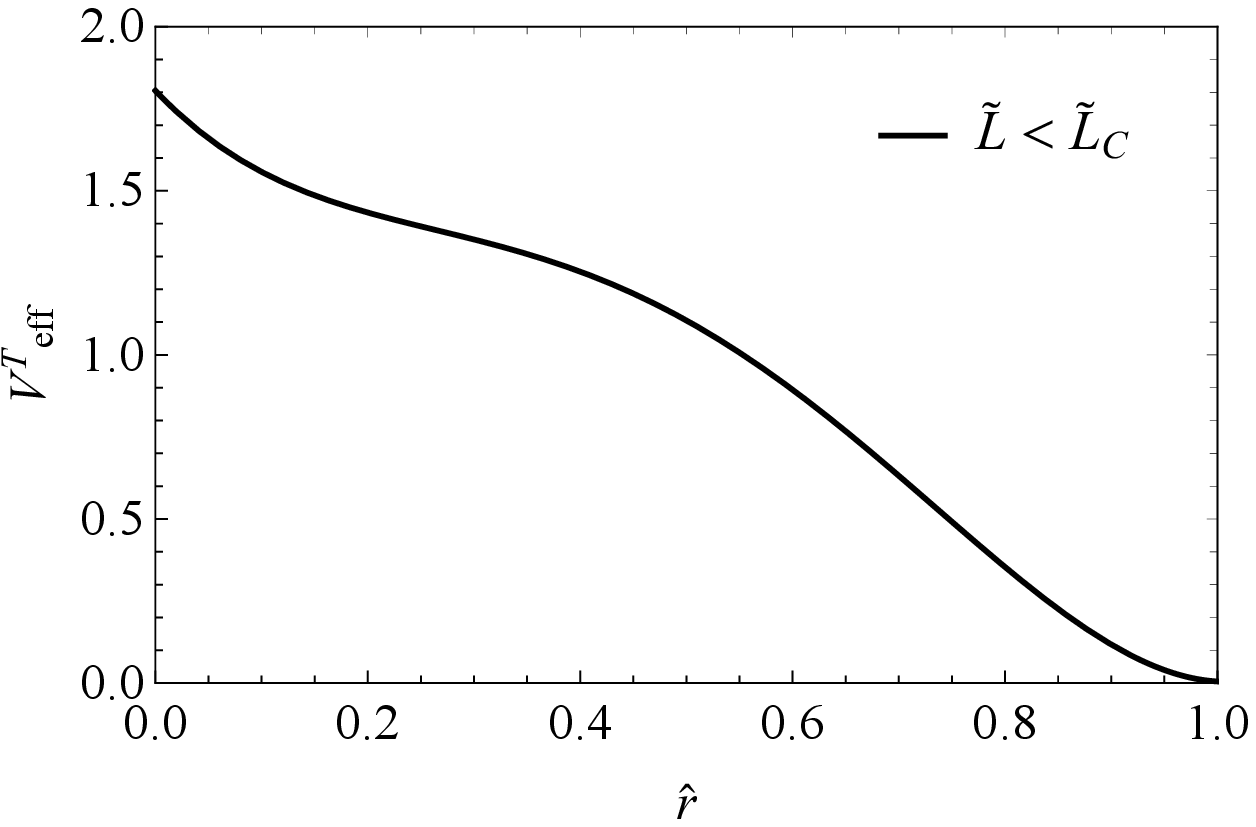}\label{ef1}}
	\subfigure[]{\includegraphics[scale=0.7]{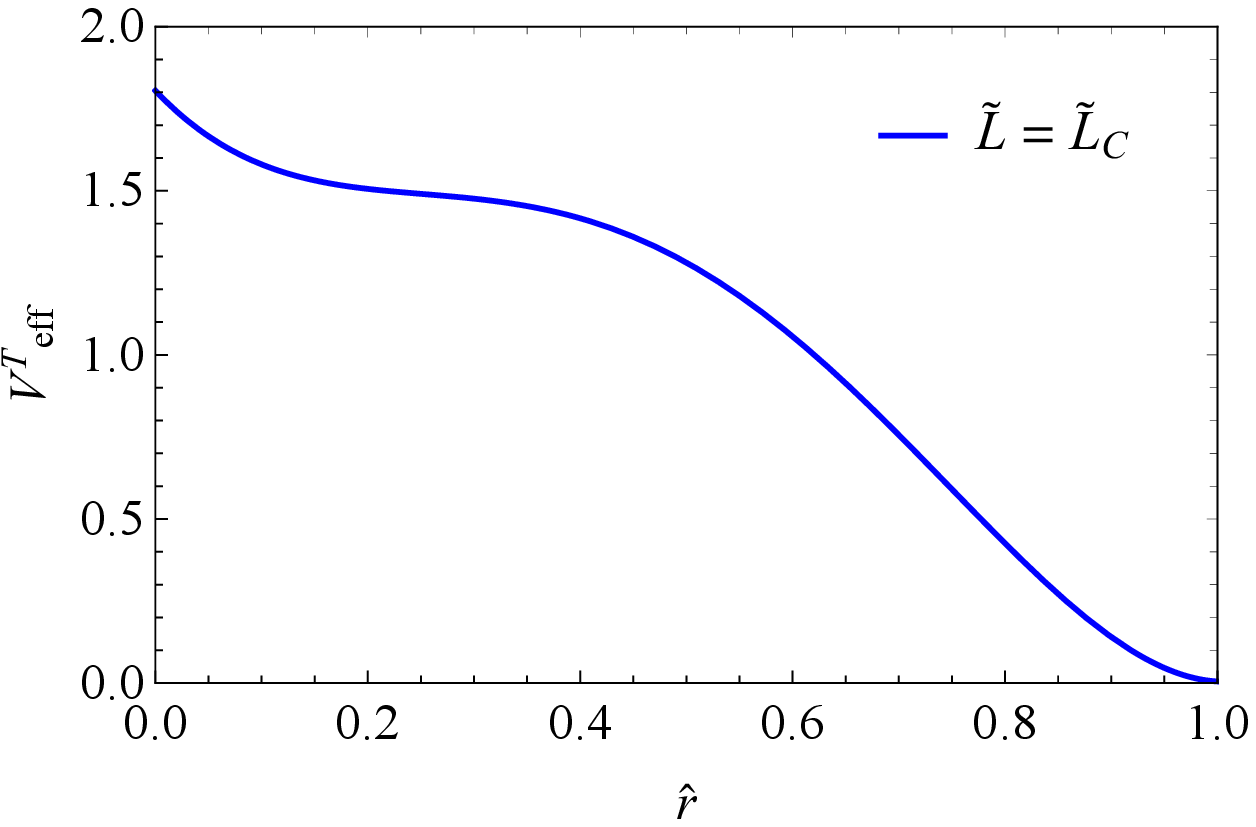}\label{ef2}}
	\subfigure[]{\includegraphics[scale=0.7]{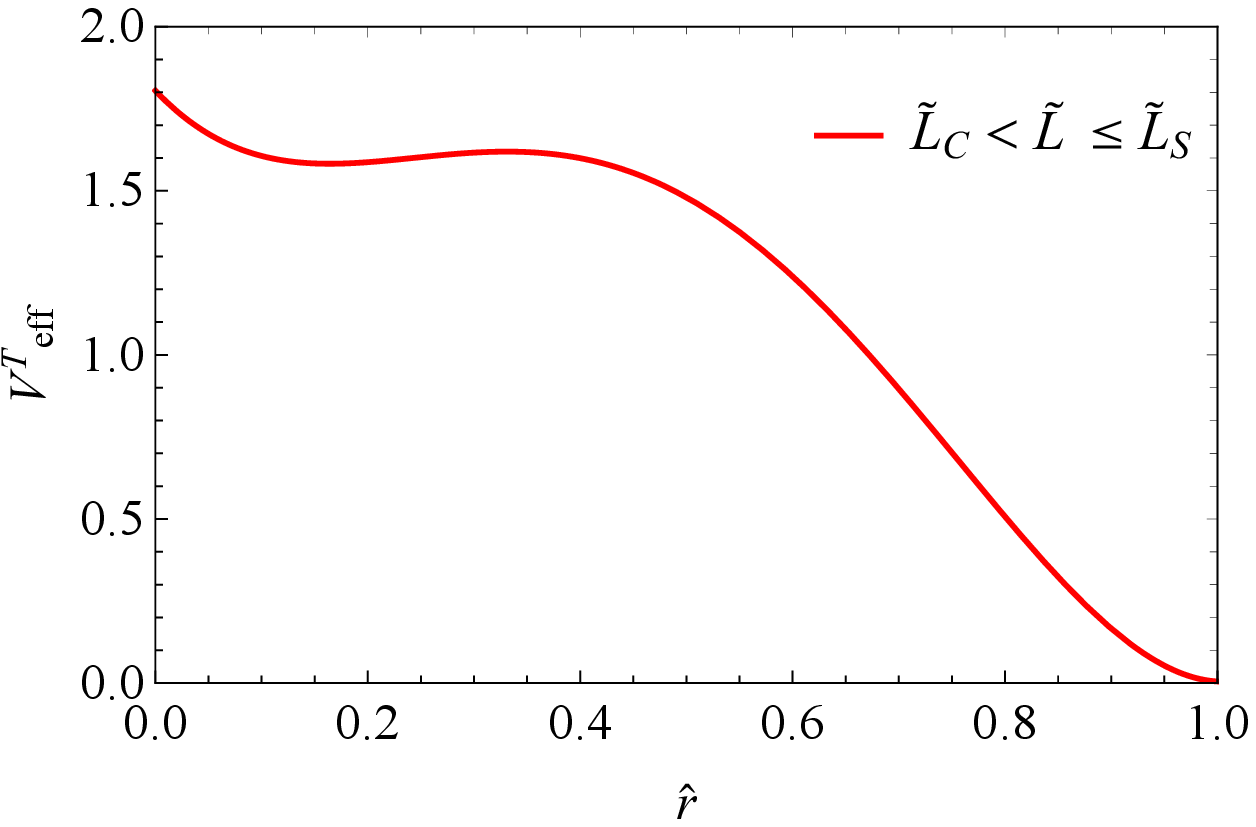}\label{ef3}}
	\subfigure[]{\includegraphics[scale=0.7]{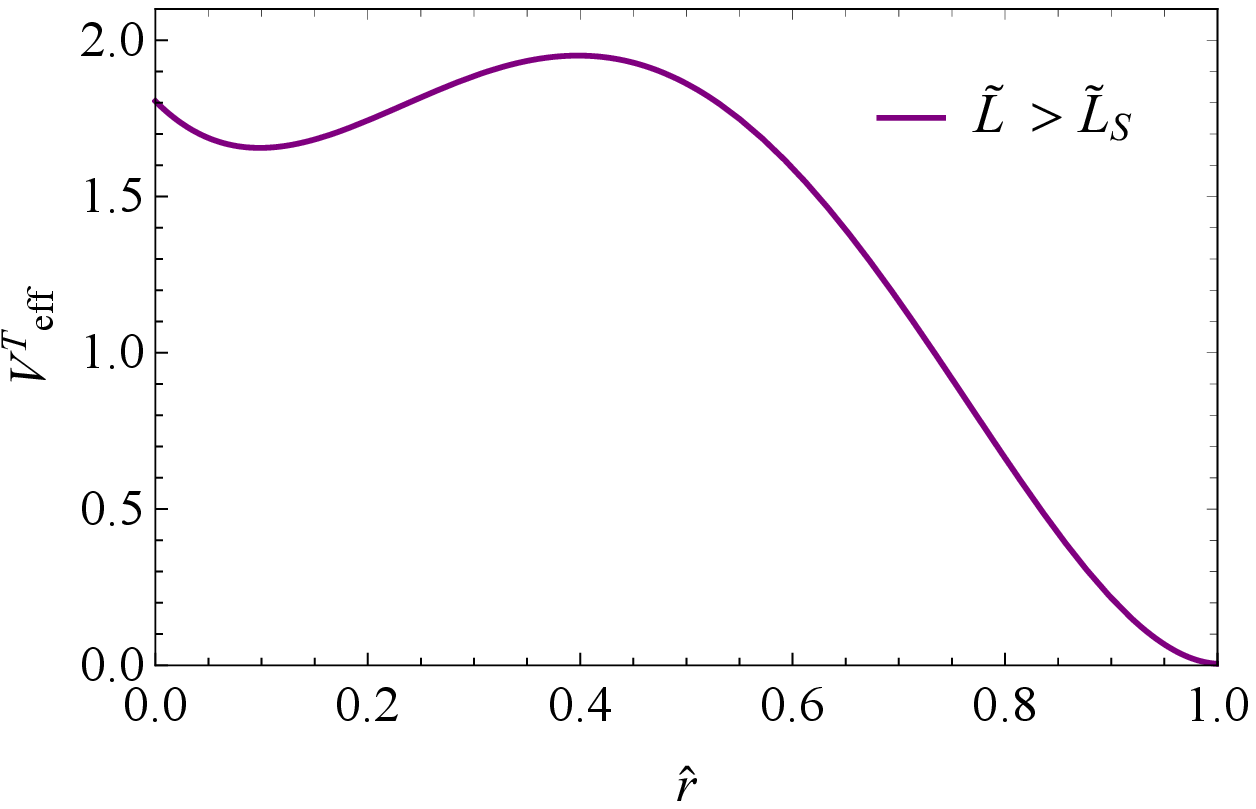}\label{ef4}}
	\caption {Variation of effective potential $V_{eff}^{T}(\hat{r})$ for timelike geodesics as a function of radial coordinate $\hat{r}$ for different values of angular momentum (a) $\tilde{L}=2.2$, (b) $\tilde{L}=2.5$, (c) $\tilde{L}=2.8$ and (d) $\tilde{L}=3.3$. Here we fix the TWH parameter $\bar{\gamma_{+}}$ to 0.9. }\label{Effectivetime}
\end{figure*}
\subsection {Lyapunov stability of timelike circular orbits } 
In this section, we analyse the stability of timelike circular geodesics. Considering $k=-1$,  Eq.\eqref{effF} leads to the effective potential for the case of timelike geodesics as
\begin{equation}
V_{eff}^{T}(\hat{r})=  2 \tilde{L}^{2} \hat{r}^{2}(1-\hat{r})^{2}+\frac{1}{2} [\bar{\gamma_{+}}(2\hat{r}-1)-1]^{2} .\label{effT}
\end{equation}
Willenborg et.al \cite{willenborg2018geodesic} has already investigated and classified the types of orbits in the background of the TWH geometry, supported by a massless conformally coupled scalar field, via illustrating the effective potential. For timelike case, depending upon the various values of the angular momentum $\tilde{L}$ and TWH parameter $\bar{\gamma_{+}}$ (see Fig.1 of ref. \cite{willenborg2018geodesic} for the dependency of characteristic values of $\tilde{L}$  on $\bar{\gamma_{+}}$), we have depicted four different types of effective potential $V_{eff}^{T}(\hat{r})$ in \figurename{\ref{Effectivetime}}. The two characteristic values of angular momentum namely critical angular momentum $\tilde{L_{C}}$ and $\tilde{L_{S}}$ which obeys $V_{eff}^{T}(\bar{\gamma_{+}},\tilde{L_{S}})> V_{eff}^{T}(\bar{\gamma_{+}},0)$ for timelike geodesics were introduced in \cite{willenborg2018geodesic}. On this basis, four different cases arise for timelike geodesics and the orbits are classified in terms of the nature of the effective potential as follows:
\vspace{0.15cm}\\
\noindent Case I: when $\tilde{L}<\tilde{L}_{C}$ i.e. $\tilde{L}=2.2$, only transit and escape orbits are found as the potential decreases monotonically. 
\vspace{0.1cm}\\
Case II: when $\tilde{L}=\tilde{L}_{C}$ i.e. $\tilde{L}=2.5$, the  unstable circular, transit and escape orbits may exist because the potential consists a double zero in allowed range of $\hat{r}$.
\vspace{0.2cm}\\
Case III: when $\tilde{L}_{C}<\tilde{L}\leq\tilde{L}_{S}$ i.e.  $\tilde{L}=2.8$, the bound orbits may also be found due to minimum of effective potential, and unstable circular orbits exist due to the maximum of effective potential.
\vspace{0.2cm}\\
Case IV: when $\tilde{L}>\tilde{L}_{S}$ i.e.  $\tilde{L}=3.3$, the bound and unstable circular orbits are found along with other orbits because the maximum of potential becomes larger than $V_{eff}^{T}(\hat{r}=0)$.\\

\begin{figure*}[]
	\centering
	\subfigure[]{\includegraphics[scale=0.6]{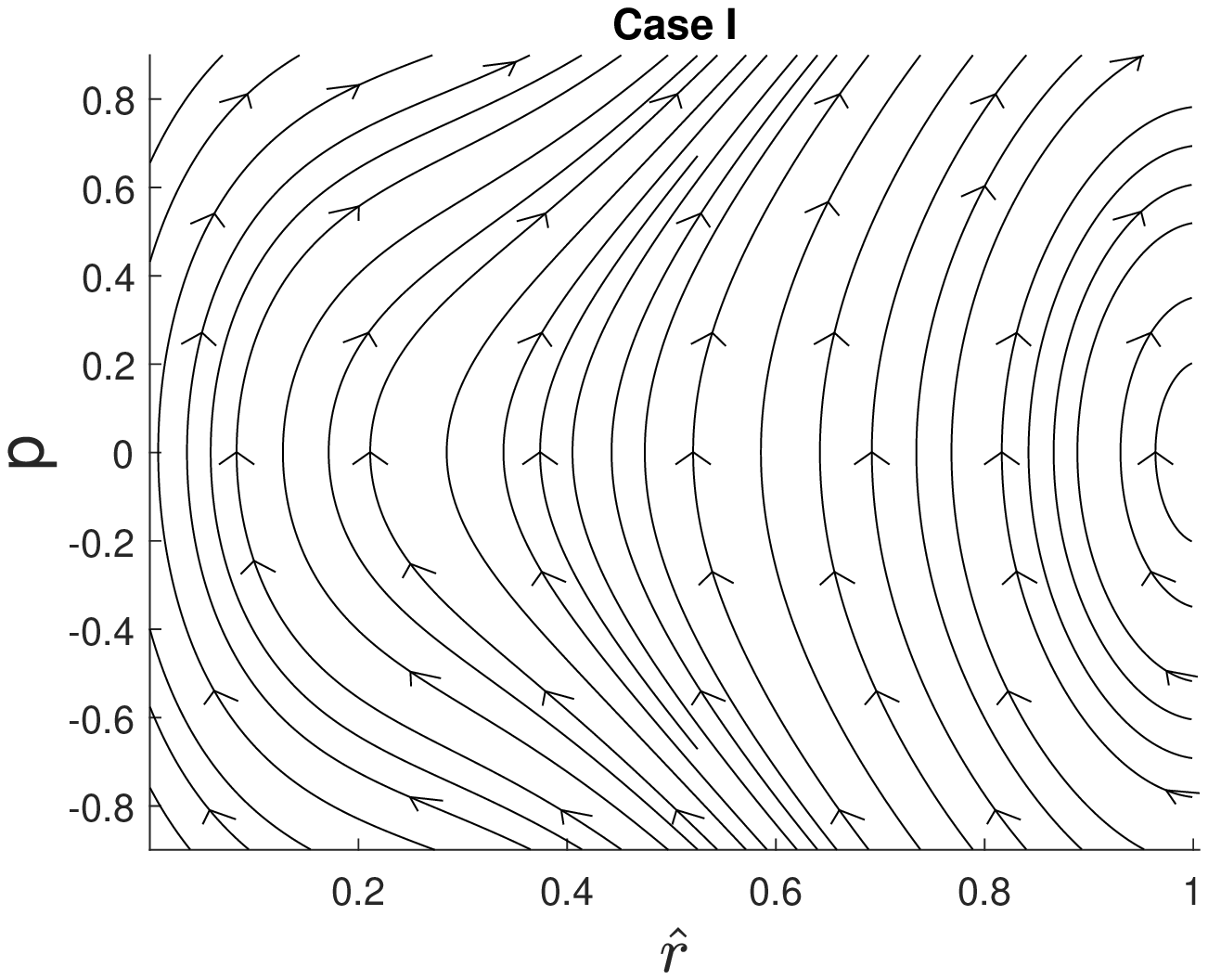}\label{ppt1}}
	\subfigure[]{\includegraphics[scale=0.6]{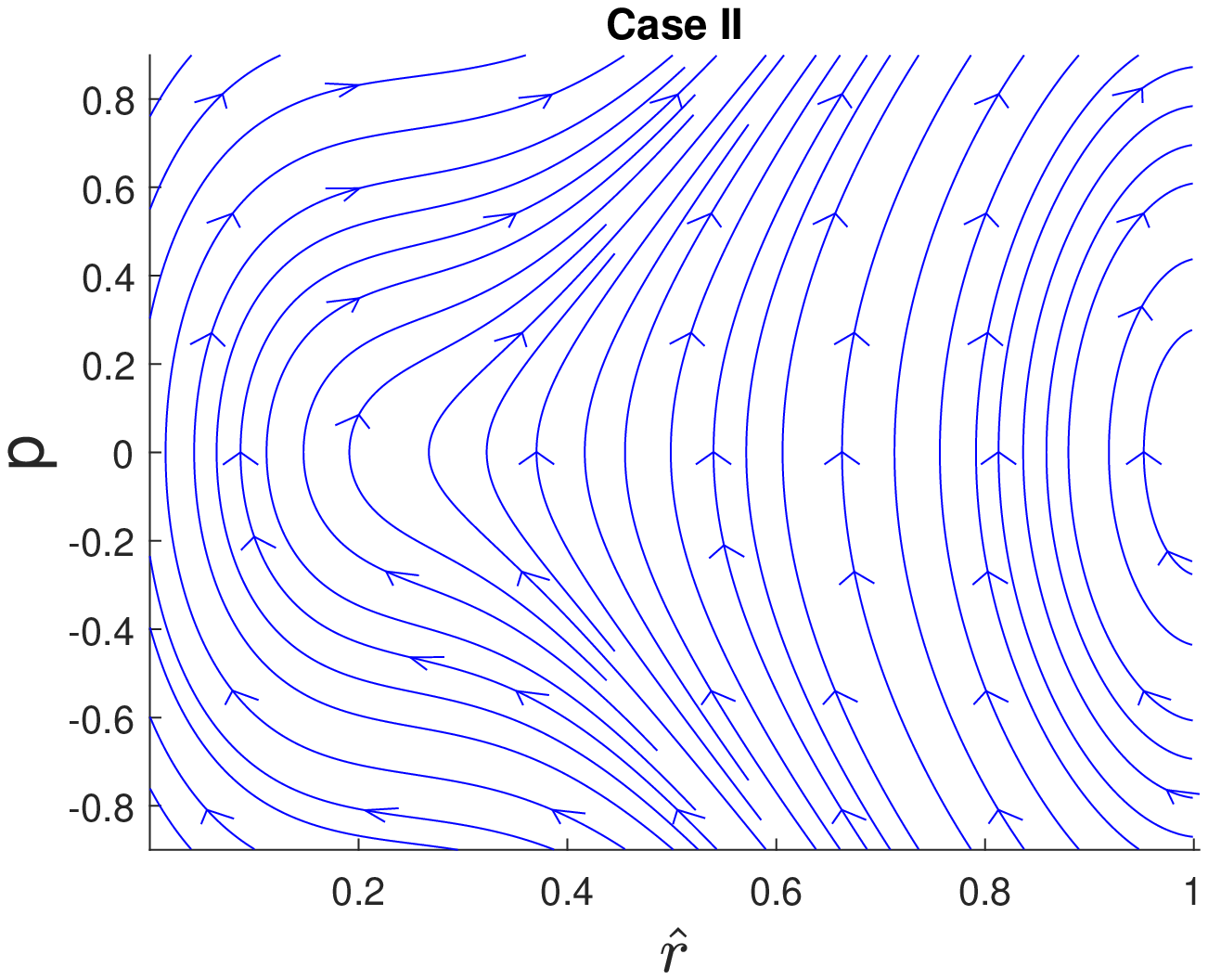}\label{ppt2}}
	\subfigure[]{\includegraphics[scale=0.6]{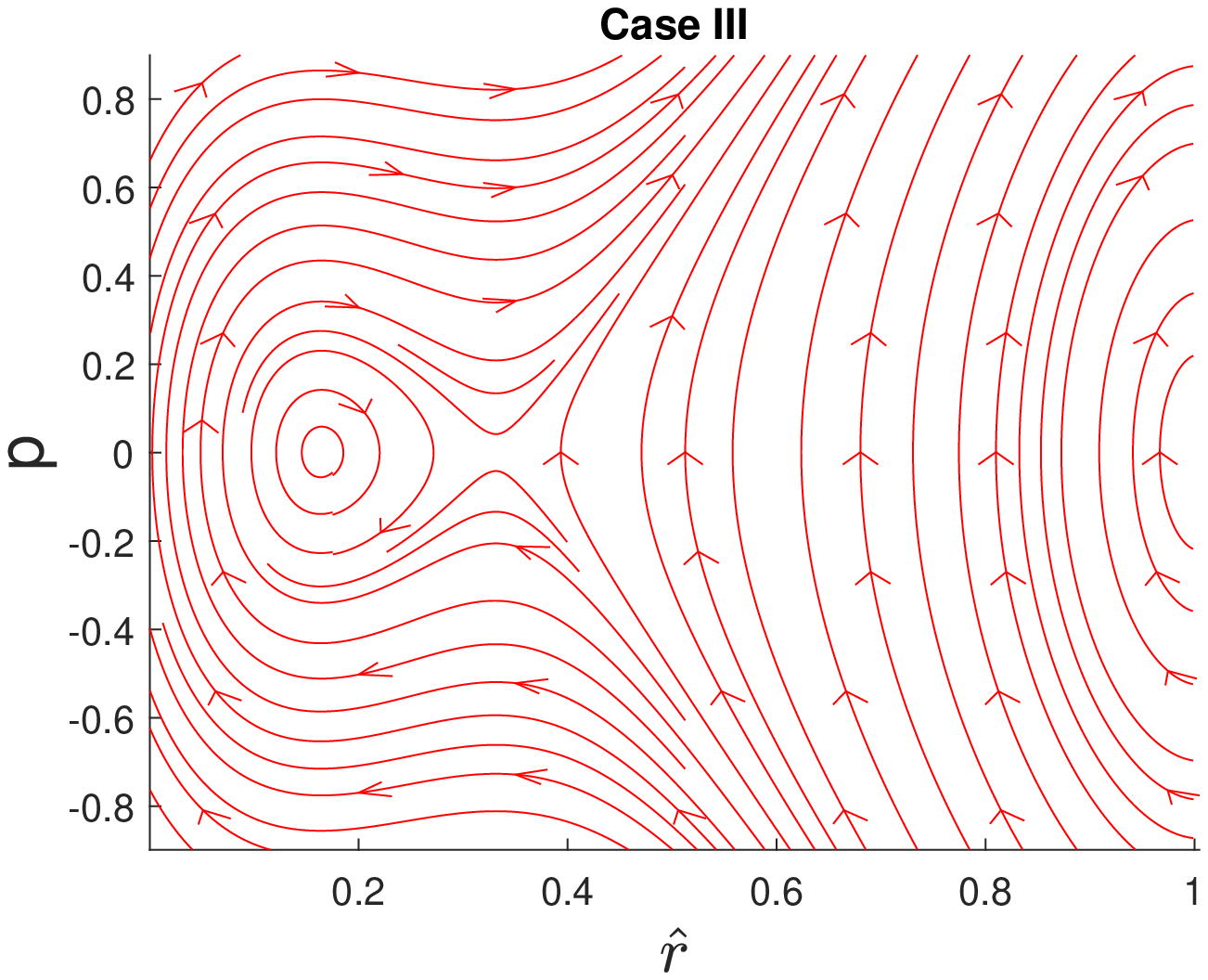}\label{ppt3}}
	\subfigure[]{\includegraphics[scale=0.6]{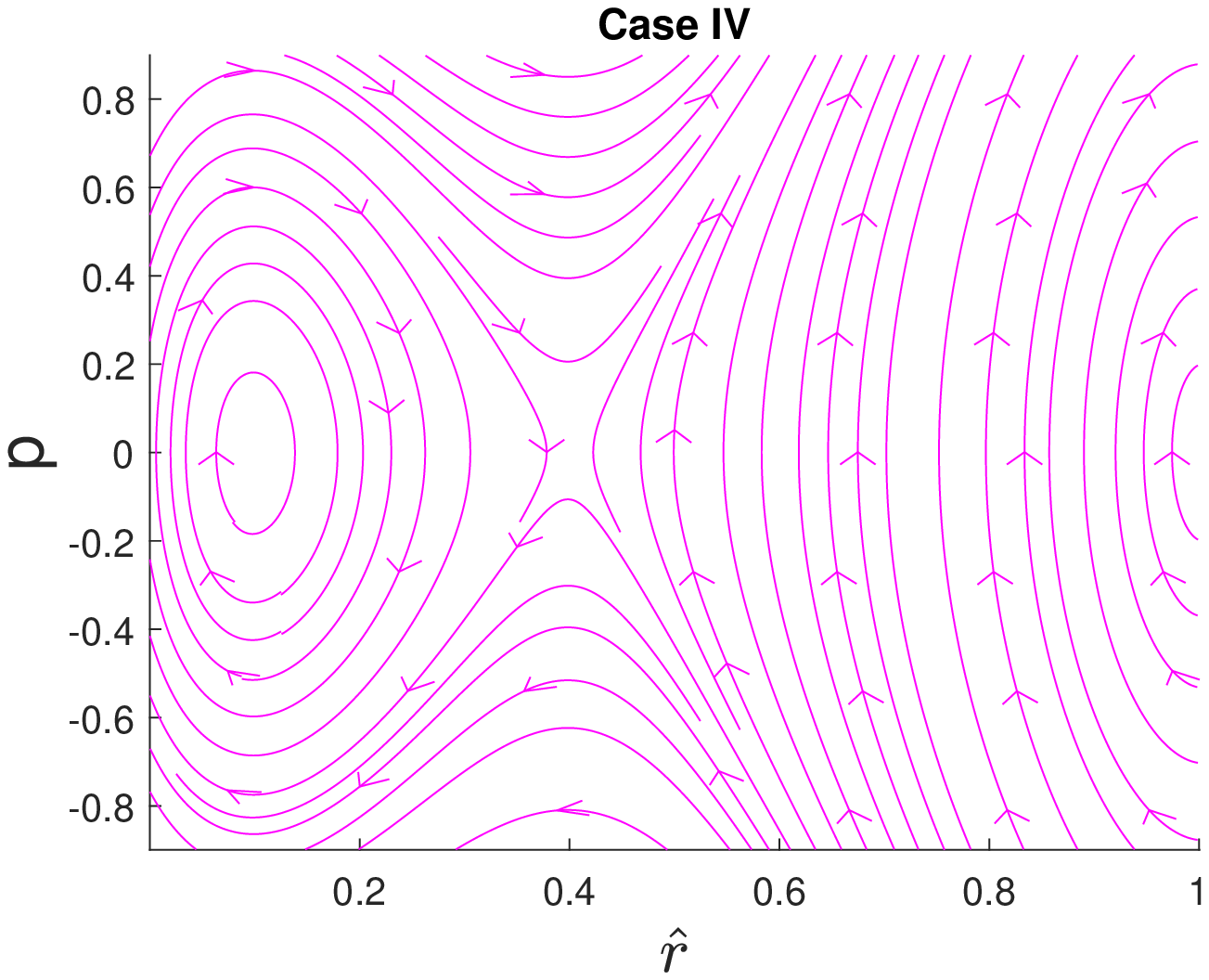}\label{ppt4}}
	\caption {The phase portrait in $\hat{r}-p$ plane for four different cases of angular momentum $\tilde{L}$ for timelike geodesics with parameter $\bar{\gamma_{+}}=0.9$. }\label{phaseporT}
\end{figure*}
\noindent To determine the fixed points of the timelike effective potential, let us take $\left(V_{eff}^{T}(\hat{r})\right)'=0$ and obtain these points shown in \tablename{ \ref{table1}}. Since in case of our spacetime metric, the radial coordinate $\hat{r}\in [0,1]$, therefore only the fixed point $(\hat{r_{0}},0)$ within this range need to be considered.
\begin{table*}[ht]
	\caption{Fixed points  ($\hat{r_{0}},0$) of $V_{eff}^{T}(\hat{r})$ for timelike geodesics for different cases with $\bar{\gamma_{+}} = 0.9$.}
	\centering
	\begin{tabular}{|c|c|c|c|}
		\hline
		$\tilde{L}<\tilde{L}_{C}$ &  $\tilde{L}=\tilde{L}_{C}$ & $\tilde{L}_{C}<\tilde{L}\leq\tilde{L}_{S}$ & 	$\tilde{L}>\tilde{L}_{S}$  \\  [0.5ex]
		\hline
		0.246096\ + 0.164556 i  & 0.246865 + 0.083853 i & 0.163923 & 0.0982812  \\
		
		0.246096\ - 0.164556 i  & 0.246865 - 0.083853 i & 0.330947 & 0.397913  \\
		
		1.00781 & 1.00627 & 1.00513 & 1.00381 \\
		\hline		
	\end{tabular}  
	\label{table1}
\end{table*}
\begin{table*}[ht]
	\caption{ Lyapunov stability of fixed points ($\hat{r_{0}},0$) for different cases in timelike geodesics.}
	\centering
	\begin{tabular}{|c|c|c|c|c|}
		\hline
		Case & Fixed point ($\hat{r_{0}},0$) & $\left(V_{eff}^{T}(\hat{r_{0}})\right)''$ & $\lambda_{T}$  & 	Lyapunov stability  \\  [0.5ex]
		\hline
		$\tilde{L}_{C}<\tilde{L}\leq\tilde{L}_{S}$	& 0.163923 & 8.81225 & Imaginary & Lyapunov stable \\
		
		& 0.330947 & -7.06259 & Real & Lyapunov unstable \\
		
		\hline
		$\tilde{L}>\tilde{L}_{S}$ &  0.0982812 & 23.6378 & Imaginary & Lyapunov stable \\
		
		&  0.397913 & -15.8162 & Real & Lyapunov unstable \\
		\hline
	\end{tabular}  
	\label{table2}
\end{table*}
\noindent The eigenvalues of the Jacobian matrix for the timelike case are given by
\begin{equation}
\lambda_{T}=\pm \sqrt{-\left(V_{eff}^{T}(\hat{r_{0}})\right)''} . 
\end{equation}
Further, by performing the Lyapunov stability criteria, the classification of fixed points ($\hat{r_{0}},0$) for each case are presented in \tablename{ \ref{table2}}. For the cases $\tilde{L}<\tilde{L}_{C}$ and $\tilde{L}=\tilde{L}_{C}$, no fixed point belongs in the range $[0,1]$ which reveals that the circular orbits (stable or unstable) do not exist in this range. However, for $\tilde{L}_{C}<\tilde{L}\leq\tilde{L}_{S}$, the fixed point $(0.163923,0)$ is Lyapunov stable as the corresponding eigenvalue is imaginary while fixed point $(0.330947,0)$ is Lyapunov unstable as the corresponding eigenvalue is real. On the other hand, for $\tilde{L}>\tilde{L}_{S}$, the fixed point $(0.0982812,0)$ is Lyapunov stable as the corresponding eigenvalue is imaginary and $(0.397913,0)$ is Lyapunov unstable as corresponding eigenvalue is real.
\vspace{0.2cm}\\
\noindent  A circular orbit exists corresponding to each fixed point of the effective potential for timelike as well as null geodesics. Furthermore, one can also differentiate the circular orbits at fixed points as saddle point (which is Lyapunov unstable) and stable center (which is Lyapunov stable) by visualizing the phase portrait. The phase portrait in $\hat{r}-p$ plane for stable and unstable timelike circular orbits are shown in \figurename{\ref{phaseporT}}. It is observed that
\vspace{0.2cm}\\
Case I ($\tilde{L}<\tilde{L}_{C}$): as shown in \figurename{\ref{ppt1}}, the stable or unstable circular orbits are not found because no fixed point exists within range $[0,1]$. 
\vspace{0.2cm}\\
Case II ($\tilde{L}=\tilde{L}_{C}$): it is clearly shown in \figurename{\ref{ppt2}}, that stable or unstable circular orbits are not found as no fixed point lies within range $[0,1]$.
\vspace{0.2cm}\\
Case III ($\tilde{L}_{C}<\tilde{L}\leq\tilde{L}_{S}$): from the corresponding phase portrait presented in \figurename{\ref{ppt3}}, we observe that the circular orbit at fixed point  $(0.163923,0)$ is a stable center, while another circular orbit at fixed point $(0.330947,0)$ is an unstable saddle point.
\vspace{0.2cm}\\
Case IV ($\tilde{L}>\tilde{L}_{S}$): from \figurename{\ref{ppt4}}, one can notice that the circular orbit at fixed point  $(0.163923,0)$ is a stable center while another circular orbit at fixed point $(0.330947,0)$ is an unstable saddle point. 

\subsection{Lyapunov stability of null circular orbits}
\begin{figure}[H]
	\centering
	\subfigure[]{\includegraphics[height=7cm,width=8.5cm]{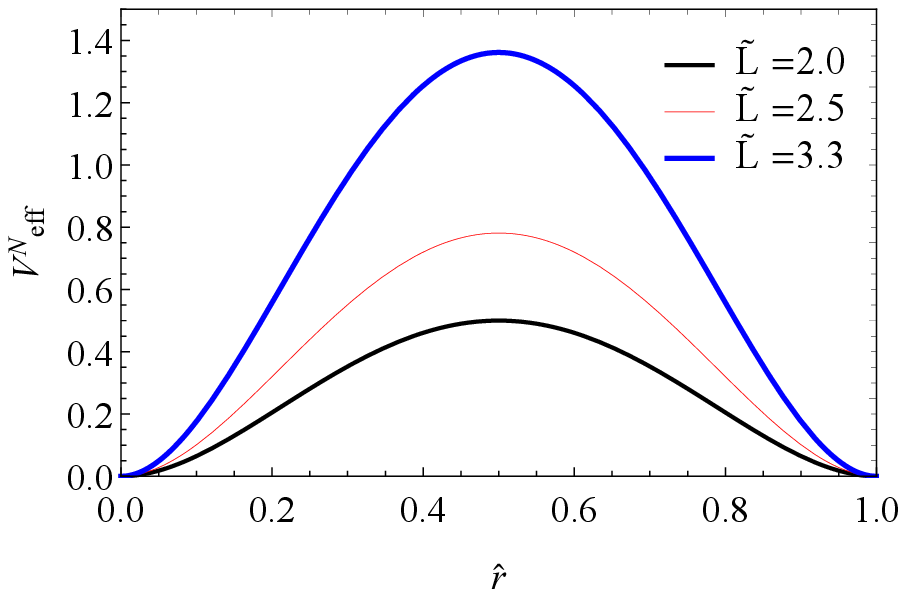}\label{effNWH}}
	\subfigure[]{\includegraphics[height=7.5cm,width=9cm]{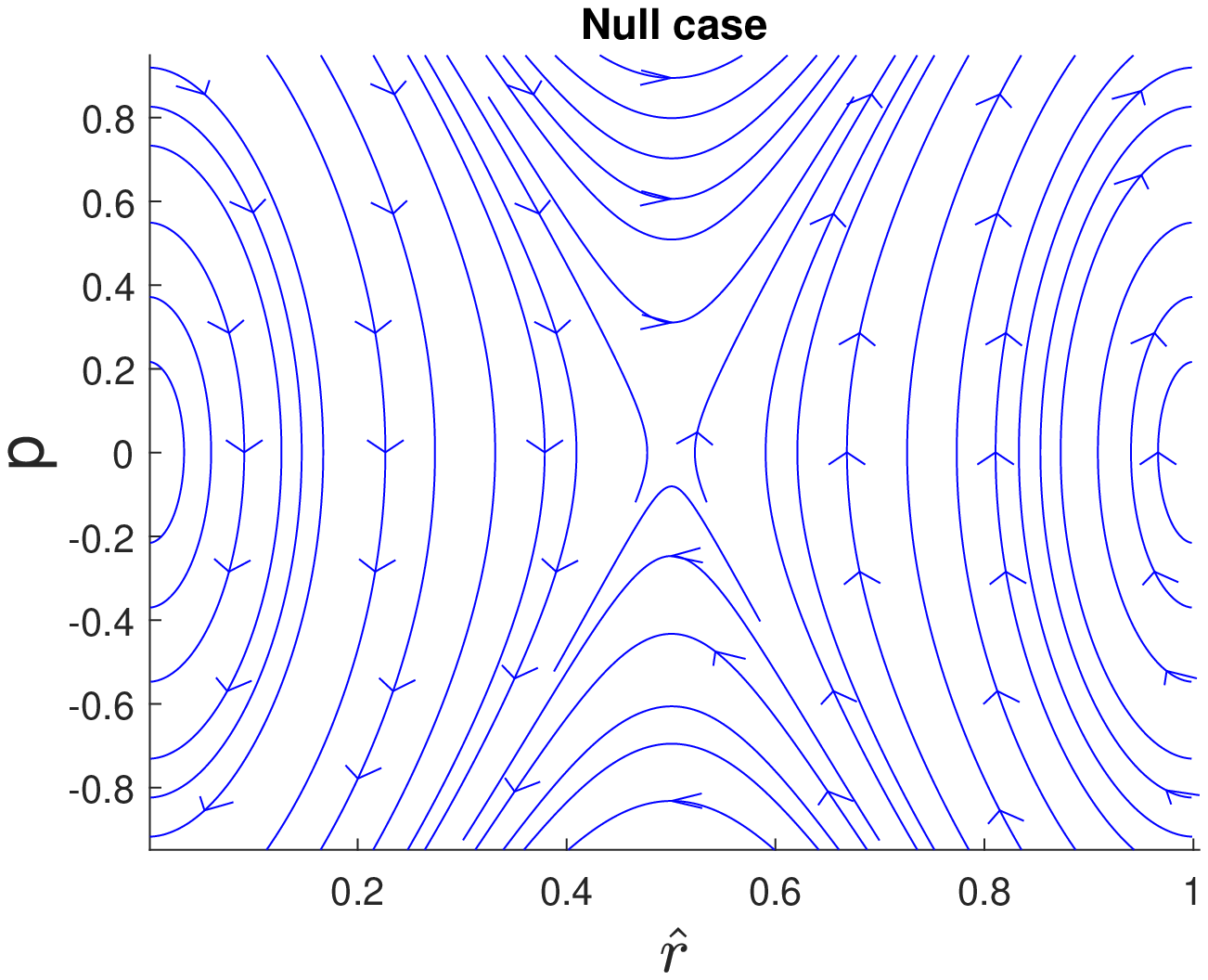}\label{ppnull}}
	\caption {(a) Variation of the effective potential $V_{eff}^{N}$ for null geodesics as a function of radial coordinate $\hat{r}$ for different values of $\tilde{L}$.  (b) The phase portrait in $\hat{r}-p$ plane with angular momentum $\tilde{L}=3.3$ for null geodesics.}
\end{figure}

From Eq.\eqref{effF}, if $k=0$ is taken into account, the effective potential for null circular geodesics that contains only the centrifugal part is expressed as
\begin{equation}
V_{eff}^{N}(\hat{r})=  2 \tilde{L}^{2} \hat{r}^{2}(1-\hat{r})^{2} .\label{effNull}
\end{equation}
\begin{table*}[ht]
	\caption{ Lyapunov stability of fixed points ($\hat{r_{*}},0$) for angular momentum $\tilde{L}=3.3$ in null geodesics.}
	\centering
	\begin{tabular}{|c|c|c|c|}
		\hline
		Fixed point ($\hat{r_{*}},0$) & $\left(V_{eff}^{N}(\hat{r_{*}})\right)''$ & $\lambda_{N}$  & 	Lyapunov stability  \\  [0.5ex]
		\hline
		(0,0) & 43.56 & Imaginary & Lyapunov stable \\
		\hline
		(0.5,0) & -21.78 & Real & Lyapunov unstable \\
		\hline
		(1,0) & 43.56 & Imaginary & Lyapunov stable \\
		\hline
	\end{tabular}  
	\label{table3}
\end{table*}
It is straight forward to note that the effective potential depends only on angular momentum $\tilde{L}$ and is independent of the parameter $\bar{\gamma_{+}}$ i.e. the effect of TWH parameter is insignificant. The behavior of the effective potential $V_{eff}^{N}$ versus $\hat{r}$ presented in \figurename{\ref{effNWH}}, shows that the height of the potential depends on the size of the angular momentum $\tilde{L}$. In the asymptotic regions i.e. $\hat{r}=0,1$, the potential $V_{eff}^{N}$ vanishes and possesses a maximum at $\hat{r}=0.5$. Hence, the types of orbits are likely to be transit and escape orbits along with unstable circular orbits for the null case. 
By using the condition $\left(V_{eff}^{N}\right)'=0$, we have determined the fixed point $(\hat{r_{*}},0) $ for null circular geodesics as $(0,0) $, $(0.5,0)$ and $(1,0)$. The Jacobian matrix for null geodesics is thus obtained giving eigenvalues corresponding to each fixed point as
\begin{equation}
\lambda_{N}=\pm \sqrt{-\left(V_{eff}^{N}(\hat{r_{*}})\right)''}. 
\end{equation}
The Lyapunov stability of each fixed point for the fixed value of angular momentum $\tilde{L}=3.3$ is visualised in \tablename{ \ref{table3}}. The fixed point $(0.5,0)$ is Lyapunov unstable as eigenvalues at this point is real while the other two points are Lyapunov stable as eigenvalues at these point are imaginary.\\
Besides, we have visualized the phase-portrait in \figurename{\ref{ppnull}} in order to decide whether the circular orbits at fixed points are stable center or a saddle point. Nevertheless, the radial coordinate $\hat{r}$ can vary from $0$ to $1$, and fixed asymptotic points $(0,0)$ and $(1,0)$ are not considered for corresponding circular orbits to be stable center.  Clearly, it is observed from the phase portrait that the circular orbit at only the fixed point $(0.5,0)$ is an unstable saddle point. Consequently, it can noticed that no stable circular orbits exist in the range $0<\hat{r}<1$ and an unstable circular orbit of photons (lightlike particle) exists only at $\hat{r}=0.5$.

\section{Conclusion} 
We have explored the Lyapunov stability analysis of circular orbits in gravitational field of a TWH geometry, conformally coupled to a massless scalar field, which is introduced by a “new improved stress-energy tensor”. Instead of the NEC violation, such WH geometry stay with GR describes gravity without any exotic type of matter distribution. Subsequently, we investigated the stability  of both particle (timelike) and photon (null) orbits in detail.  Based on our study, we have differentiated the fixed points according to their Lyapunov stability of effective potential for both timelike and null geodesics. For timelike geodesics, on classifying the the four different types of effective potentials (in view of conditions to angular momentum $\tilde{L}$), there are no fixed points obtained in the range  of radial coordinate $0\leq\hat{r}\leq1$ for the cases $\tilde{L}<\tilde{L}_{C}$ and $\tilde{L}=\tilde{L}_{C}$.  However, for $\tilde{L}_{C}<\tilde{L}\leq\tilde{L}_{S}$, we found $(0.163923,0)$ as a Lyapunov stable point  and $(0.0.330947,0)$ as a Lyapunov unstable point. Also in the case of $\tilde{L}>\tilde{L}_{S}$, the fixed points $(0.0982812,0)$ and $(0.397913,0)$ are found to be  Lyapunov stable and unstable respectively. Indeed, each fixed point of the effective potential is referred as a circular orbit of a particle about the origin. Further, by visualizing the corresponding $(\hat{r}-p)$ phase-portrait, we observed that the timelike circular orbit at every fixed point is either a stable center or  an unstable saddle point (visualized in the corresponding phase-space diagrams). Turning to lightlike particles (photons), by deducing the effective potential, the fixed points $(0,0)$ and $(1,0)$ are identified as Lyapunov stable while only one point $(0.5,0)$ is found to be Lyapunov unstable. The phase-portrait regarding null geodesics also revealed that the null circular orbit at the Lyapunov unstable fixed point (i.e. $0.5,0$) is obtained as a saddle point. As a consequence, we note that stable circular orbits do not exist for the case of null geodesics. 
\section*{\normalsize Acknowledgments}
{\normalsize S.G. gratefully acknowledges University Grants Commission (UGC), New Delhi, India for providing the financial support  as a Senior Research Fellow through UGC-Ref.No. {\bf 1479/CSIR-UGC NET-JUNE2017}. The authors also acknowledge the facilities available at ICARD, Gurukula Kangri (Deemed to be University) Haridwar those were used during the course of this work.  }  

\bibliographystyle{unsrt} 
\bibliography{shobitrefTWH} 
\end{document}